\documentclass{elsart}

 \usepackage{graphicx}
 \usepackage{epsfig}

\usepackage{amssymb}

\begin{document}

\begin{frontmatter}


 \title{Bonabeau hierarchy models revisited}
 \author{Lucas Lacasa}
 \ead{lucas@dmae.upm.es}

 \address{Departamento de Matem{\'a}tica Aplicada y Estad{\'i}stica\\
ETSI Aeron{\'a}uticos\\ Universidad Polit{\'e}cnica de Madrid.}

\title{}

\author{Bartolo Luque}
\ead{bartolo@dmae.upm.es}
\address{Departamento de Matem{\'a}tica Aplicada y Estad{\'i}stica\\
ETSI Aeron{\'a}uticos\\ Universidad Polit{\'e}cnica de Madrid.}

\begin{abstract}
\noindent What basic processes generate hierarchy in a collective?
The Bonabeau model provides us a simple mechanism based on
randomness which develops self-organization through both
winner/looser effects and relaxation process. A phase transition
between egalitarian and hierarchic states has been found both
analytically and numerically in previous works. In this paper we
present a different approach: by means of a discrete scheme we
develop a mean field approximation that not only reproduces the
phase transition but also allows us to characterize the complexity
of hierarchic phase. In the same philosophy, we study a new version
of the Bonabeau model, developed by Stauffer et al. Several previous
works described numerically the presence of a similar phase
transition in this later version. We find surprising  results in
this model that can be interpreted properly as the non-existence of
phase transition in this version of Bonabeau model, but a changing
in fixed point structure.
\end{abstract}

\end{frontmatter}

\section{Introduction}

\noindent  It is usual, in sociological works, to describe how
global behavior appears, in many levels of social activities
\cite{libro}. Before that, it is more fundamental to understand in
which way citizens gather \cite{Sociologia1}: since in a little
collective every single seems to play the same status, in big
societies diversity appears \cite{Stauffer6}. Hierarchic dominance
and hierarchic stratification has been studied with several
different approaches \cite{Sociologia1,Sociologia2,Sociologia3}. As
long as these matters can be considered as many-body dynamical
systems, they have attracted the attention of physicists in the
latest years. The emergent area of Sociophysics involves those
social complex systems, dealing with many different social
situations with a statistical
physics approach \cite{AXELROD,libro,Stauffer6}.\\
In this terms, a simple and fruitful model of diversity generation
has been proposed by Bonabeau et al. \cite{libroBonabeau,Bonabeau1}.
Related to this model, some modified versions have been proposed, as
the Stauffer et al. \cite{Stauffer1} version, or the
one from Ben-Naim and Redner \cite{REDNER} (this later one has been solved analytically).\\
The purpose of this paper is double: first, by means of a discrete
mean field approximation, we reproduce analytically the numerical
results found by Bonabeau. We discover a non trivial complex
structure in the hierarchy generation path. After this, we apply the
same scheme to Stauffer version \cite{Stauffer1,Stauffer2}, widely
used as a model of hierarchy generation
\cite{Stauffer1,Stauffer2,Stauffer3,Stauffer4,Stauffer5,Gallos}, in
order to obtain analytical evidences of its numerical behavior.
\section{The Bonabeau model}
\noindent  The Bonabeau model \cite{libroBonabeau,Bonabeau1} has
been proposed as a simple model showing self-organization to explain
hierarchic dominance in Ethology. With subtle modifications it has
been reallocated in Sociophysics area as a model of social
stratification \cite{Stauffer2,Stauffer1,Stauffer5,REDNER}. It
starts from a community composed of  $N$ agents, randomly
distributed over a regular lattice $L \times L$, that is to say,
with a population density $\rho = \frac{N}{L\times L}$. Each agent
$i=1,2,...,N$ is characterized by a time dependent variable
$h_i(t)$, the agent fitness, that from now on we will name status.
Initially all agents share the same status $h_{i}(t=0)=0$: the
so-called egalitarian
situation. System dynamics are:\\\\
(1) \emph{Competition with feedback}: an agent $i$ chosen randomly
moves in a four nearest neighbor regular lattice (Newmann
neighborhood). If the target site is empty, the agent takes the
place. If it is already occupied by an agent $j$, a fight occurs.
The attacker agent $i$ will defeat agent $j$ with some probability:

\begin {equation}
P_{ij}(t)=\frac{1}{1 + \exp\big(\eta(h_{j}(t)-h_{i}(t))\big)}.
\label{probabilidades}
\end {equation}
Where $\eta>0$ is a free parameter. If $i$ wins, he exchanges
positions with $j$. Otherwise, positions are maintained. After each
combat, status $h_i(t)$ are updated, increasing by $1$ the winner's
status and decreasing by $F$ the looser's status. Note that $F$ is a
parameter of the system that weights the defeats such that $F\geq1$.
The case $F=1$ will be the symmetric case from now on. In the
asymmetric case, $F>1$, the fact of loosing will be more
significative for the individual status than the fact of winning
 \cite{Stauffer2}.\\\\
(2) \emph{Relaxation}: a natural (Monte Carlo) time step is defined
as $N$ movement processes (with or without combat). After each time
step all agents update their status by a relaxation factor
$(1-\mu)$, such that $0 < \mu < 1$; this effect is interpreted
as a fading memory of agents.\\

Notice that competition rule (1) is a feedback mechanism: status
differences $h_{j}(t)-h_{i}(t)$ drive the future winning/loosing
probabilities of agents $i$ and $j$. If agent $i$ wins/looses it's
winning/loosing probability increases afterwards. This mechanism
amplifies agent inhomogeneity. On the other hand, relaxation rule
(2) drives the agent status $h_{i}(t)$  to equalize: status
differences are absorbed and toned town. The balance between both
mechanisms generates asymptotic stability on $h_{i}(t)$. Common
sense would lead us to expect low fights when the agent's density is
low, so that the relaxation mechanism would overcome and egalitarian
situation would prevail ($h_{i} = h_{j} = 0,\forall i,j$). But if
the system possesses high agent's density, the rate of fights would
increase, and competition mechanism will prevail, leading the system
to non neglecting inhomogeneities. The balance of these two
mechanisms is crucial at a given density $\rho$. Simulations ran by
Bonabeau et al.\cite{libroBonabeau,Bonabeau1} show how this
compromise between both effect bring about a phase transition at a
critical density, between egalitarian societies for low densities
and hierarchical societies for high densities.
\newline

A natural measure for the status diversification is the standard
deviation of its stationary distribution $\{h^*_i\}_{i=1,...,N}$.
However, in \cite{Stauffer2} another measure is proposed: this is
the standard deviation of stationary probability distribution
$\{P_{ij}\}_{i=1,...,N}$, defined as:

\begin{equation}
\sigma =\big(\langle P_{ij}^2 \rangle - \langle P_{ij} \rangle ^2
\big)^\frac{1}{2}. \label{sigma}
\end{equation}

\noindent This choice turns out to be more suitable , as long as it
is a bounded parameter: $\sigma \in [0,1]$. It works as an order
parameter of the system: for densities lower than the critical,
every $h_i$ are equal and therefore all $P_{ij}$ too, then $\sigma =
0$. For densities bigger than the critical, status are different and
therefore probabilities are also different: this leads
to a non zero value of $\sigma$.\\

\section{Mean field approximation in the Bonabeau model}

\noindent In order to tackle the system in a mathematical way, we
will obviate spatial correlations, reinterpreting $\rho$ as the
probability of two agents combat. In the spatial correlated model
this is equivalent to a random mixing of the agent's positions every
time step. Therefore, at each time step, an agent $i$ will
possess:\\
(1) Probability $1-\rho$ of no combat. In that case, the agent will
only suffer relaxation.\\
(2) Probability $\rho$ of leading a combat (with probability
$1/(N-1)$ the attacked agent will be $j$). In that case: agent $i$
will increase, in average, its status by one with probability
$P_{ij}(t)$, and will decrease by $F$ its status with probability
$1-P_{ij}(t)$. Relaxation will also be applied in this case.

\noindent The model is then described by an $N$ equation system
($i=1,...,N$) of the following shape:

\small
\begin{equation}
 h_i(t+1)= (1-\rho)(1-\mu)h_i(t) + \frac{\rho(1-\mu)}{N-1}
\sum_{j=1;j \neq i}^N \bigg\{P_{ij}(t)\big(h_i(t)+1\big)+
\big(1-P_{ij}(t)\big)\big(h_i(t)-F\big)\bigg\}. \label{Necuaciones}
\end{equation}
\normalsize

\noindent In order to analyze the system let's start with the
simplest version $N=2$. Having in mind that for two agents
$P_{12}(t)= 1-P_{21}(t)$, the system (\ref{Necuaciones}) will reduce
to:
\begin{eqnarray}
&&h_1(t+1)=(1-\mu)h_1(t) + \rho(1-\mu)\big\{P_{12}(t)(1+F)-F \big\}\nonumber \\ \nonumber \\
&&h_2(t+1)=(1-\mu)h_2(t) + \rho(1-\mu)\big\{1-P_{12}(t)(1+F)\big\}.
 \label{dosagentes}
\end{eqnarray}

\noindent In the egalitarian phase (below the critical density), the
fixed point $(h_1^*, h_2^*)$ of the two-automata system
\ref{dosagentes} will have stationary status of the same value, say
$h_1^*=h_2^*$. In order to find the fixed point of the system we can
define the mean status of the system as: $\langle h(t) \rangle =
(h_1(t)+h_2(t))/2$. The system turns into:
$$
\langle h(t+1)\rangle = (1-\mu)\langle h(t)\rangle +
\frac{\rho(1-\mu)(1-F)}{2} ,$$ \noindent with a fixed point:

\begin{equation}
\langle h^* \rangle = \frac{\rho(1-\mu)(1-F)}{2\mu} \label{hmedio2},
\end{equation}

\noindent always stable (this can be interpreted as the system's energy, which is conserved).\\

\noindent Notice that, in the egalitarian phase (below critical
density), we have $h_1^*=h_2^*=\langle h^* \rangle$. In order to
check out stability of $(\langle h^* \rangle,\langle h^* \rangle)$
we compute the Jacobian matrix of the system, evaluated in that
fixed point:

\begin{equation}
J = (1-\mu)\left( \begin{array}{cc}
                  1-A & A \\
                  A   & 1-A
                  \end{array}\right),
\end{equation}
where:
\begin{equation}
A= -\frac{\rho\eta(1+F)}{4}.
\end{equation}
\noindent By stability analysis we conclude that egalitarian phase
is stable as long as:
\begin{equation}
\rho<\rho_c=\frac{2\mu}{\eta(1-\mu)(1+F)}. \label{rho_2}
\end{equation}

\begin{figure}[t]
\centering
\includegraphics[width=0.45\textwidth]{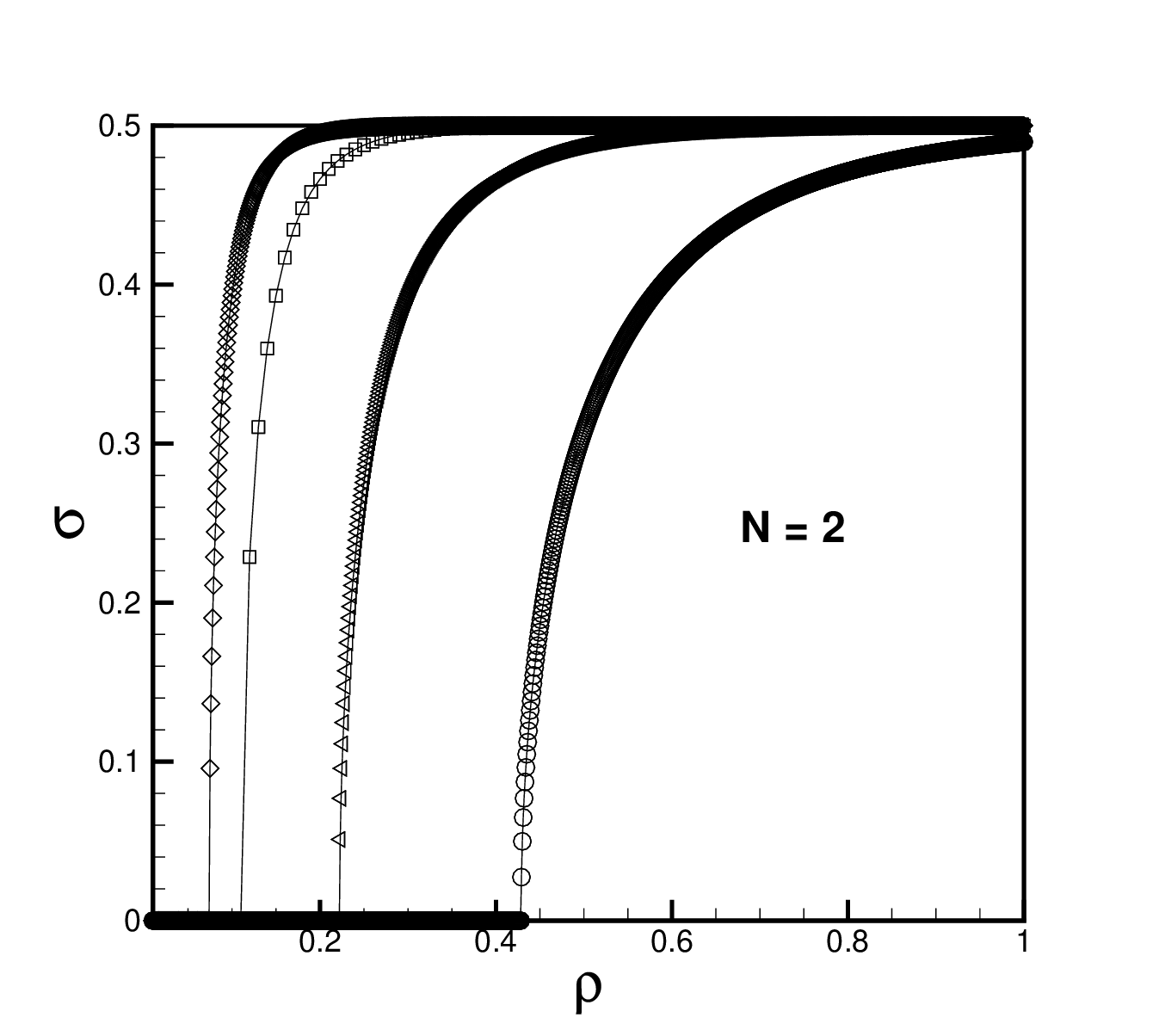}
\caption{Equality-hierarchy phase transition with control parameter
$\rho$ and order parameter $\sigma$, in the mean field model of two
automata ($N=2$). From left to right we represent the transition
derived from iteration of the two automata equation system
\ref{dosagentes}, for different values $(F;\mu;\eta)$: $(2;0.1;1)$
diamonds, $(1;0.1;1)$ squares, $(1;0.1;0.5)$ left triangles, y
$(1;0.3;1)$ circles. The continuous lines are just guides for eye.}
\label{transicionN2}
\end{figure}

\noindent From numerical iteration of the equations for
two-automata, we present in figure \ref{transicionN2} control
parameter $\rho$ versus order parameter $\sigma$, at stationary
situation, for different values of parameters $F$, $\eta$ and $\mu$.
Critical density values $\rho_c$ agree  with  (\ref{rho_2}). For the
symmetric case ($F=1$) with $\eta=1$ and $\mu=0.1$ (squares) we
obtain a critical density $\rho_c \approx 0.11$. We'll take this
particular case as the reference case from now on.

\begin{figure}[t]
\centering
\includegraphics[width=0.45\textwidth]{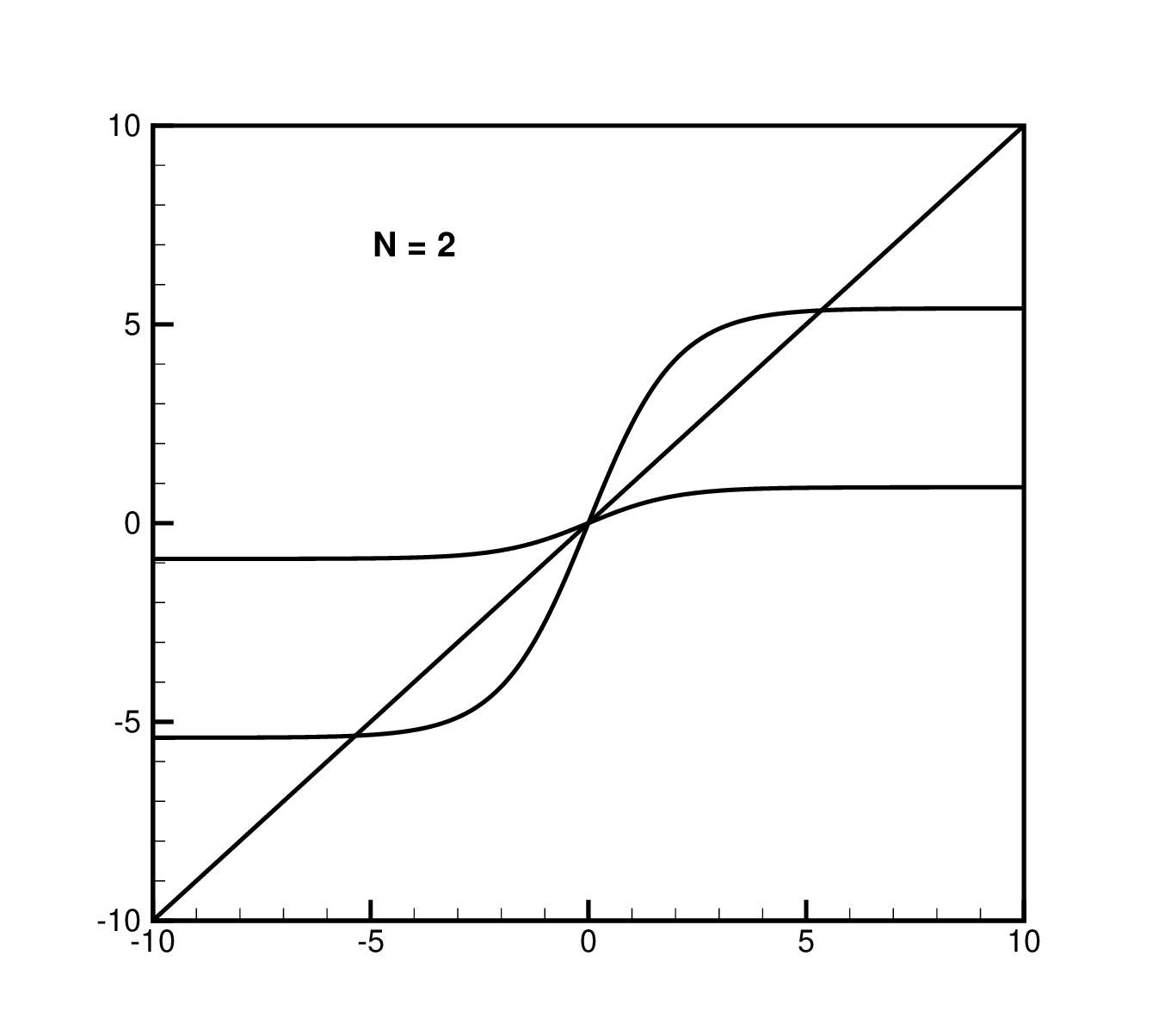}
\caption{The two automata system switches from having one fixed
point $h^*=h_2^*-h_1^*=0$ for densities  $\rho<\rho_c$, to three
fixed points when $\rho>\rho_c$ : $h^*=\{0,+a,-a\}$ (first one
unstable and the rest stable),  equivalent to the three fixed points
of the two automata system, say
$\{(0,0),(+a/2,-a/2),(-a/2,+a/2)\}.$} \label{pfijos_MAYOOO}
\end{figure}

\begin{figure}[t]
\centering
\includegraphics[width=0.45\textwidth]{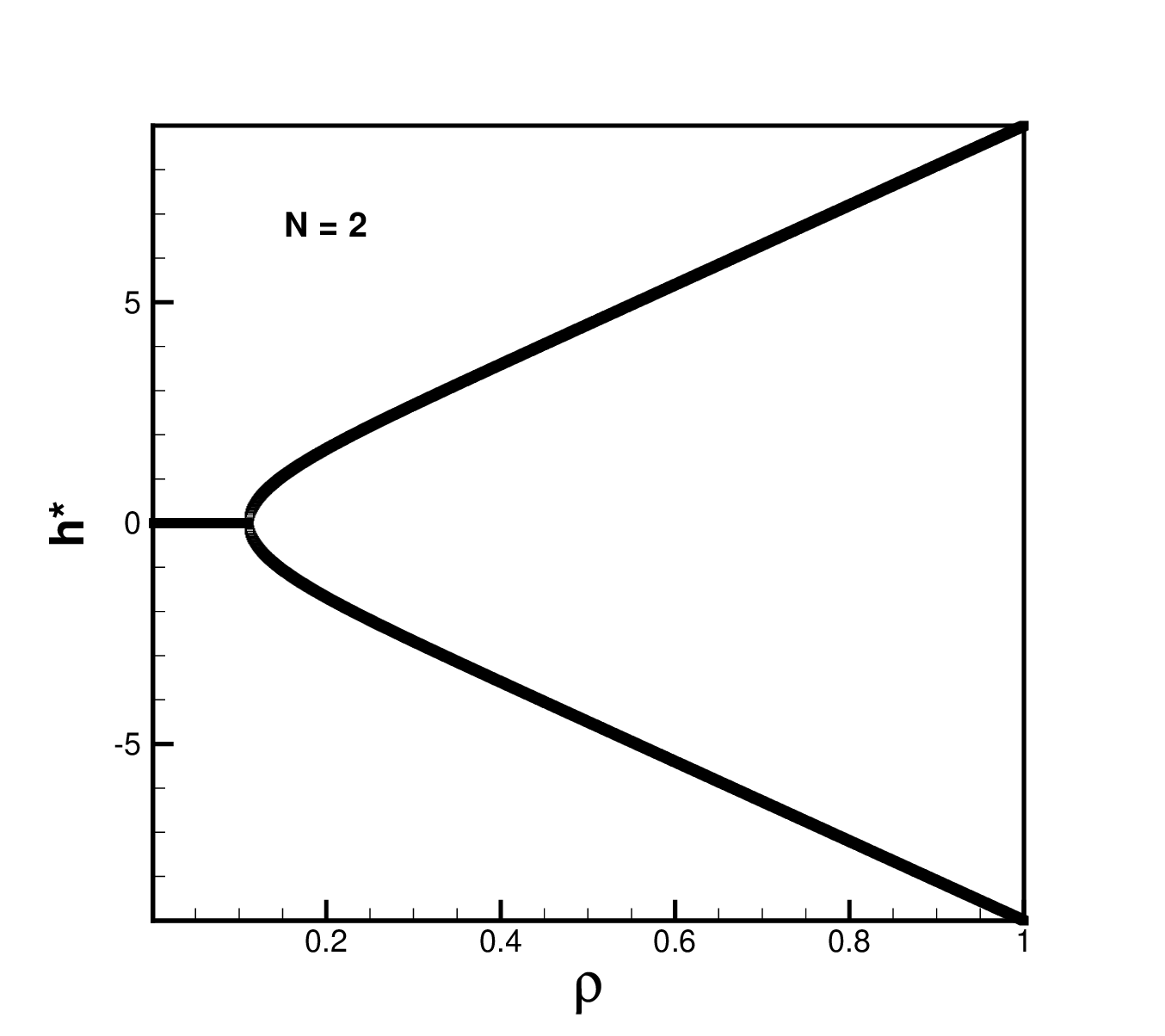}
\caption{Bifurcation of stationary values $h_1^*$ and $h_2^*$ of the
reference case (symmetric case $F=1$ with $\eta=1$ and $\mu=0.1$).
At the egalitarian zone we have: $h_1^*=h_2^*=\langle h^*
\rangle=0$. At $\rho=\rho_c$ a bifurcation occurs, from where, due
to symmetry, $h_1^*=-h_2^* \neq 0$.} \label{N2}
\end{figure}

\begin{figure}[t]
\centering
\includegraphics[width=0.45\textwidth]{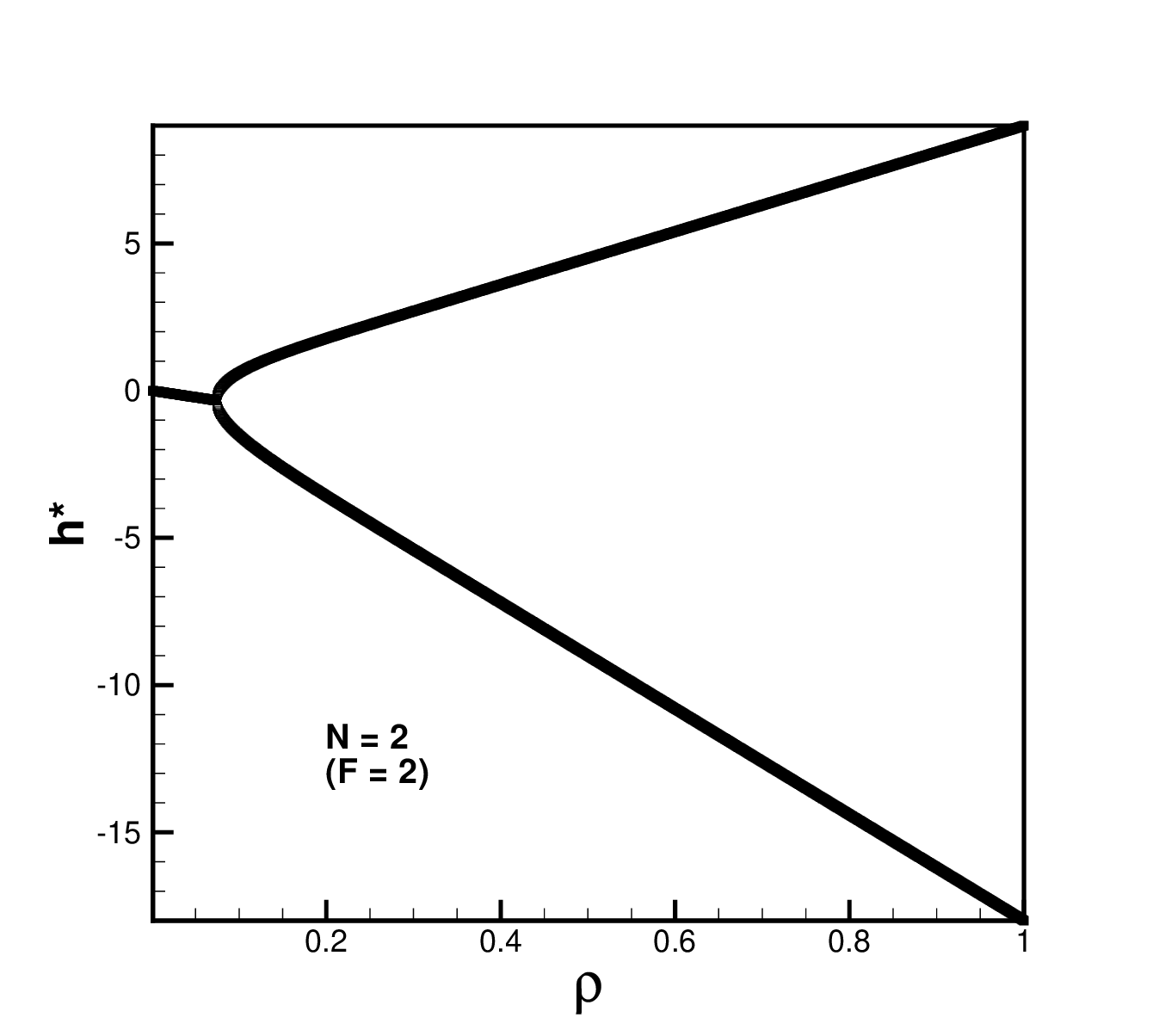}
\caption{Bifurcation of stationary values of  $h_1^*$ and $h_2^*$
for values $F=2$, $\mu=0.1$ and $\eta=1.0$.} \label{N2_F2}
\end{figure}

\begin{figure}[t]
\centering
\includegraphics[width=0.45\textwidth]{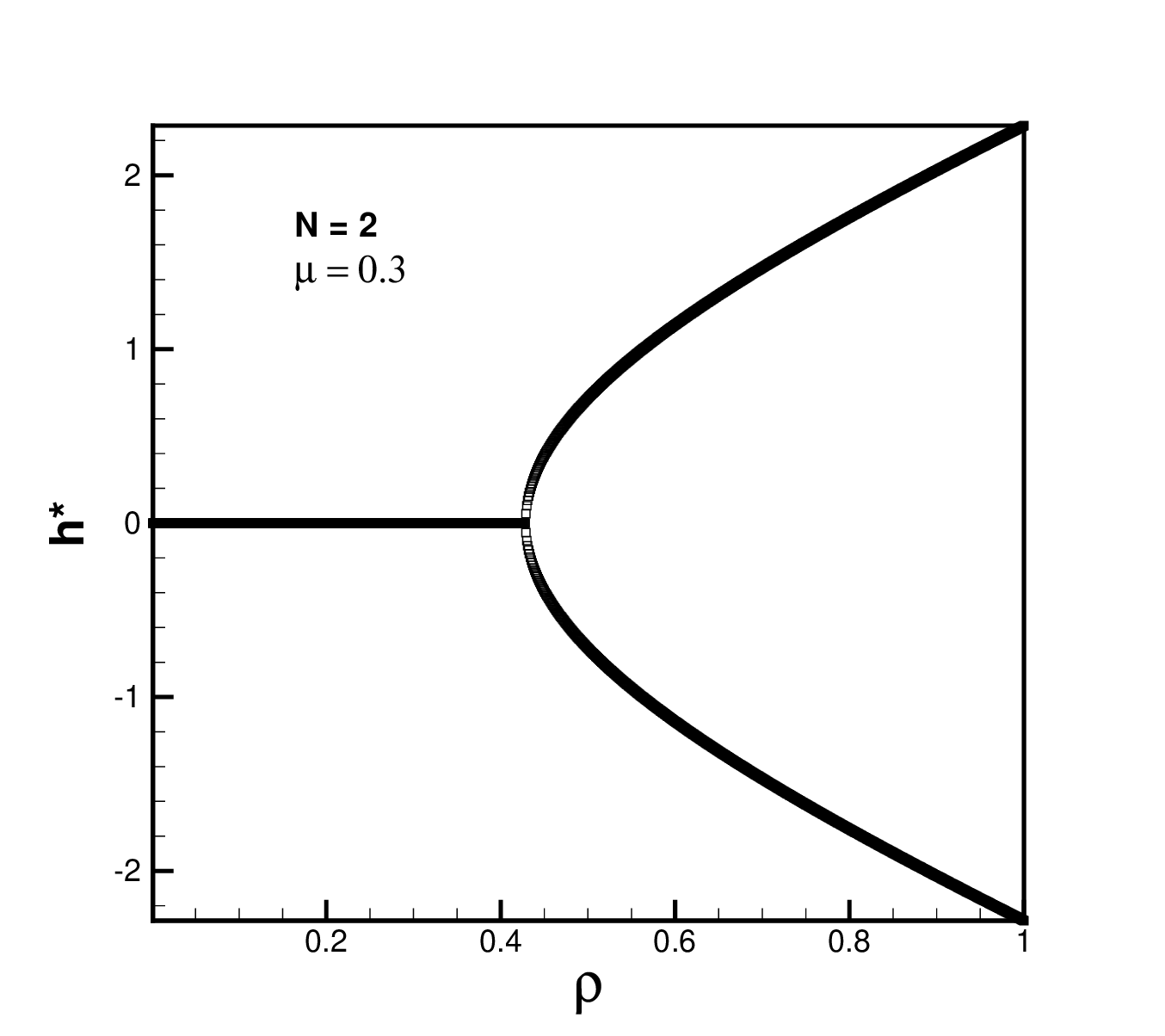}
\caption{Bifurcation of stationary values of  $h_1^*$ and $h_2^*$
for values $F=1$, $\mu=0.3$ and $\eta=1.0$.} \label{N2_mu03}
\end{figure}

\begin{figure}[t]
\centering
\includegraphics[width=0.45\textwidth]{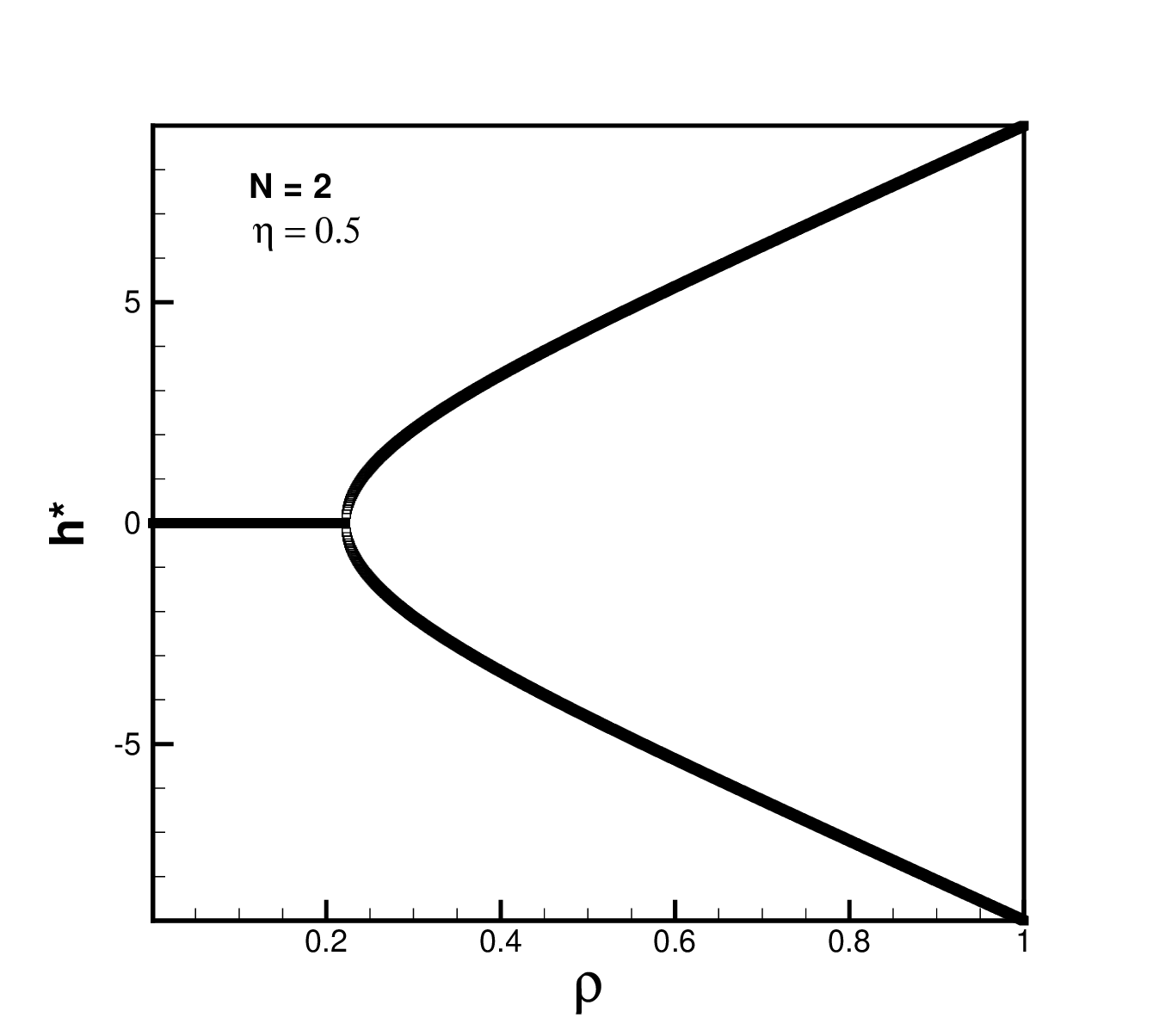}
\caption{Bifurcation of stationary values of  $h_1^*$ and $h_2^*$
for values $F=1$, $\mu=0.1$ and $\eta=0.5$.} \label{N2_eta_05}
\end{figure}

\noindent In order to understand the dynamics of the fixed points of
the two-automata system, we apply the following  change of
variables: $h^*=h_2^*-h_1^*$. The fixed points of the system become
now the solutions $h^*$ of:
\begin{displaymath}
h^*=\frac{\rho(1-\mu)(1+F)}{\mu}\bigg(1-\frac{2}{1+\exp(\eta
h^*)}\bigg).
\end{displaymath}
\noindent As to the case of reference, notice that for each value of
$\rho$ the system, as a fact of symmetry, has two fixed points:
${(h_1^*,-h_1^*) , (h_2^*,-h_1^*)}$. In figure (\ref{pfijos_MAYOOO})
we represent a case below transition ($\rho<\rho_c$), with a
singular solution, and a case above transition ($\rho>\rho_c$) with
three  solutions: this results in the solution bifurcation of
$h_1^*$ and $h_2^*$ in  figure \ref{N2}. In the egalitarian phase we
have: $h_1^*=h_2^*=\langle h^* \rangle=0$. At $\rho=\rho_c$ a
bifurcation occurs, from which we have $h_1^*=-h_2^* \neq 0$.
\newline
In the successive figures we can observe what effects produce a
variation of parameters $F$, $\eta$ and $\mu$ on status
stratification, as for the reference case (figure \ref{N2}): (1)
Increasing the asymmetry $F$ (figure \ref{N2_F2}) decreases $\rho_c$
and grows up inequality. (2) Increasing the relaxation $\mu$ (figure
\ref{N2_mu03}) increases $\rho_c$ and diminish inequality. (3) A
decrease of $\eta$ (figure \ref{N2_eta_05}) increases $rho_c$ and
has no effect on inequality.\\\\

\noindent We now apply the same philosophy to the general system
\ref{Necuaciones}. We may define the mean status of the system as:
$\langle h(t) \rangle=\frac{1}{N} \sum_{i=1}^{N}h_i(t)$. Now due to
the fact that $P_{ij}+P_{ji}=1$ and then
$\sum_{i=1}^{N}\sum_{j=1,j\neq i }^{N}P_{ij}=\frac{N(N-1)}{2}$, we
can sum  and normalize the $N$ equations, obtaining:

$$\langle h(t+1) \rangle = (1-\mu)\langle h(t) \rangle -
\frac{\rho(1-\mu)(F-1)}{2},$$ \noindent whose fixed point is
independent of the number of automata and agrees with the very first
result given at (\ref{hmedio2}) in the case of two-automata (the
system's energy). Again we get that $(h_1^*, h_2^*,\dots,h_N^*)$
with $h_1^*=h_2^*=\dots =h_N^*=\langle h^* \rangle$ is a fixed point
of the system whose stability determines the transition. The
Jacobian matrix of the linearized system, evaluated at the fixed
point, is:

$$
J=(1-\mu)\left(
            \begin{array}{cccc}
              1-A & \frac{A}{N-1} & ...& \frac{A}{N-1} \\
              \frac{A}{N-1} & 1-A & ...&... \\
              ... & ... & ... & ... \\
              \frac{A}{N-1}&...&...&1-A\\
            \end{array}
          \right),
$$
\\\noindent a circulating matrix \cite{circulante} where
$A=\frac{-\rho\eta(1+F)}{4}$ and its eigenvalues being:
$\lambda=(1-\mu)((1-A)-\frac{A}{N-1})$ with multiplicity $N-1$, and
$\lambda=(1-\mu)$ with multiplicity 1. The egalitarian phase is
therefore determined by:
$$\rho<\rho_c=\frac{4\mu(N-1)}{\eta(1-\mu) N(1+F)}.$$
\noindent This result is on agreement with the particular case of
two-automata ($N=2$), and if $N>>1$ we have:
\begin{equation}
\rho_c=\frac{4\mu}{\eta(1-\mu)(1+F)}.
\end{equation}
\noindent Notice that as long as $0 \le \rho_c \le 1$, the phase
transition will only occur under:
\begin{equation}
\mu<\frac{\eta(1+F)}{4+\eta(1+F)}.
\end{equation}

\begin{figure}[t]
\centering
\includegraphics[width=0.45\textwidth]{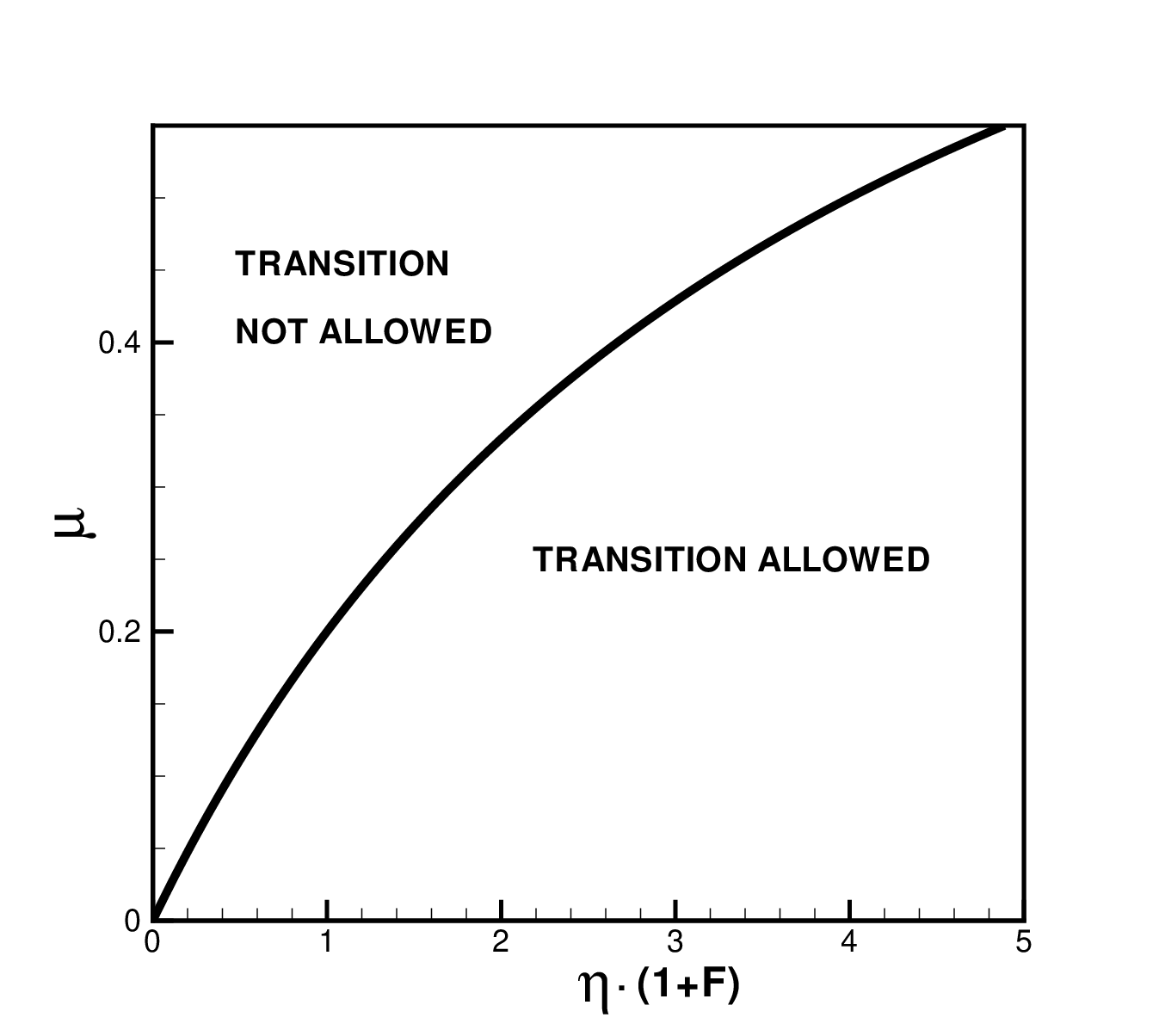}
\caption{Parameter space: delimitation, for $N>>1$, of the regions
where transitions can be wether achieved or not achieved.}
\label{fases}
\end{figure}

 In  figure (\ref{fases}) we represent, for $N>>1$, the
parameter space, where we distinguish the zone where the
egalitarian-hierarchical transition is allowed.

\section{Additive relaxation}

Bonabeau et al. in their seminal paper \cite{Bonabeau1}, proposed an
additive relaxation as an alternative to the multiplicative
relaxation developed above. That additive relaxation updates the
status as it follows:
\begin{equation} h_i\longrightarrow h_i-\mu\tanh(h_i).
\end{equation}
They developed then a mean field approximation, based on stochastic
differential equations, where they found the
egalitarian-hierarchical phase transition. We can of course apply,
in the discrete model that we propose, this additive mechanism of
relaxation. In the two-automata system, mean field equations reduce
to:
$$h_1(t+1)=h_1(t) + \rho(P_{12}(t)(1+F)-F)-\mu\tanh(h_1)$$
$$h_2(t+1)=h_2(t) + \rho(1-P_{12}(t)(1+F))-\mu\tanh(h_2).$$

\noindent Following our previous steps, it's quite easy to deduce
that the fixed point $h^*_1=h^*_2$ is stable as long as
$\rho<\rho_c=\frac{2\mu}{\eta(1+F)}$. This result is on agreement
 with those of the continuum model proposed by Bonabeau et al.
\cite{Bonabeau1}.

A simplified model has been recently proposed \cite{REDNER} by
Ben-Naim and Redner, in order to obtain analytical evidences of the
phase transition which is observed in the Bonabeau model. In their
version (which is highly inspired on the Bonabeau model) the
relaxation process is additive (though it is not a function of the
status but simply a fixed constant $h_i\longrightarrow h_i-1$), and
the competition mechanism is not stochastic but deterministic
(except for the situation where both agents have the same status). A
phase transition between both regimes (equality/hierarchy) is then
found analytically.

\begin{figure}[t]
\centering
\includegraphics[width=0.45\textwidth]{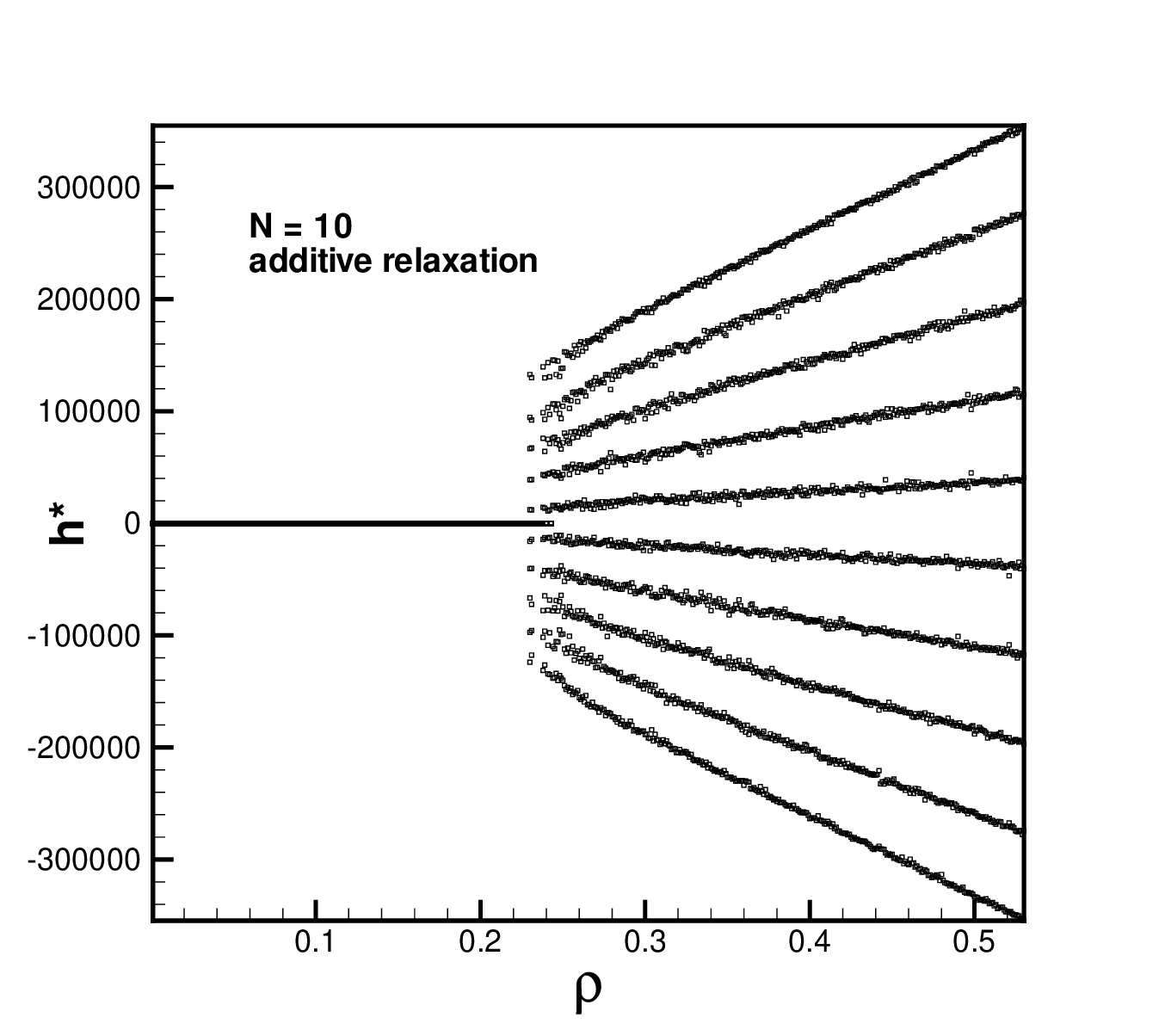}
\caption{Bifurcations of stationary values of fixed point components
(system \ref{Necuaciones}) depending on $\rho$, for the symmetric
case of reference, in the additive relaxation model, when $N=10$.}
\label{additive}
\end{figure}

\begin{figure}[t]
\centering
\includegraphics[width=0.45\textwidth]{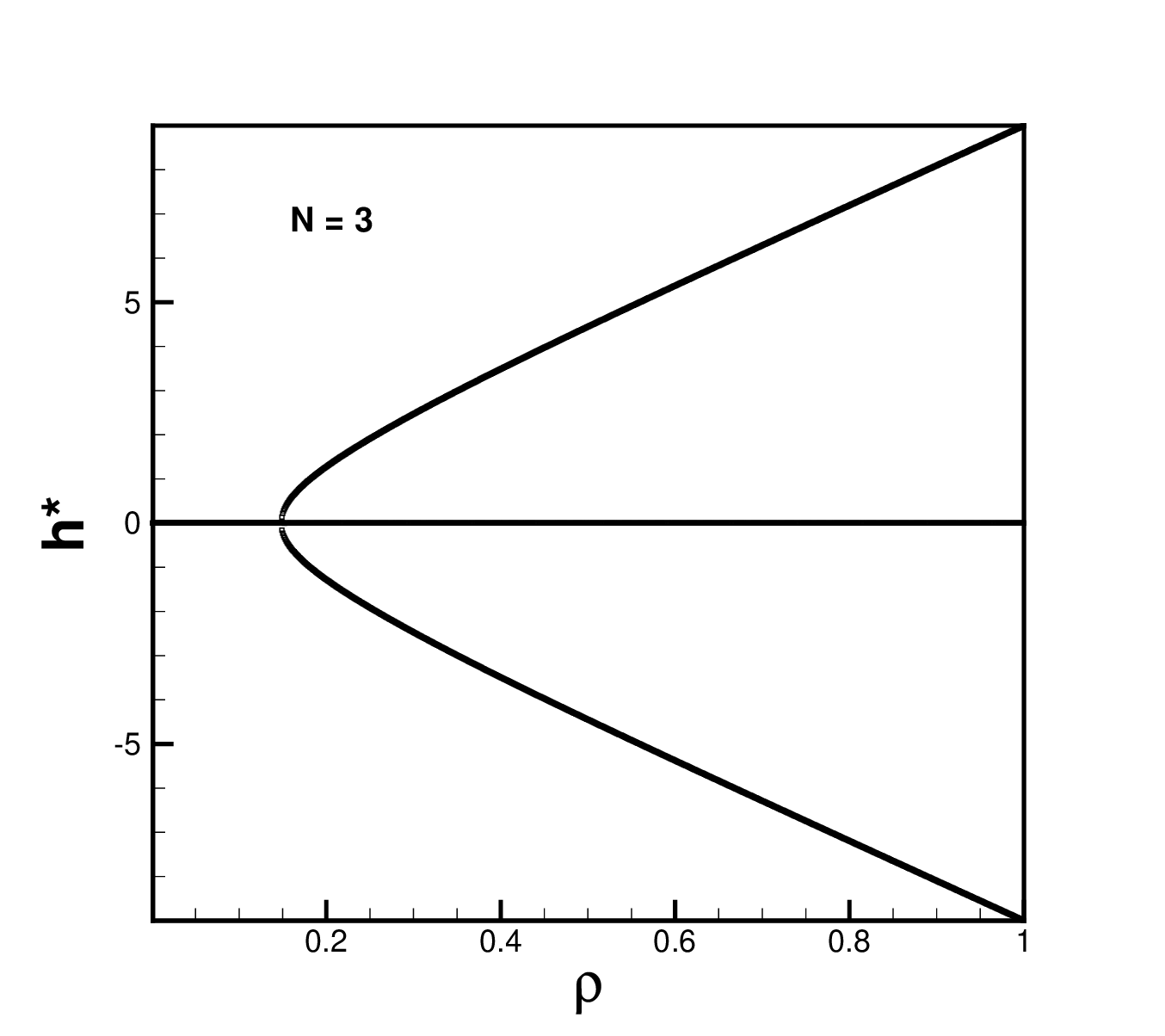}
\caption{Bifurcations of stationary values of fixed point components
(system \ref{Necuaciones}) depending on $\rho$, for the symmetric
case of reference when $N=3$.} \label{N3}
\end{figure}

\begin{figure}[t]
\centering
\includegraphics[width=0.45\textwidth]{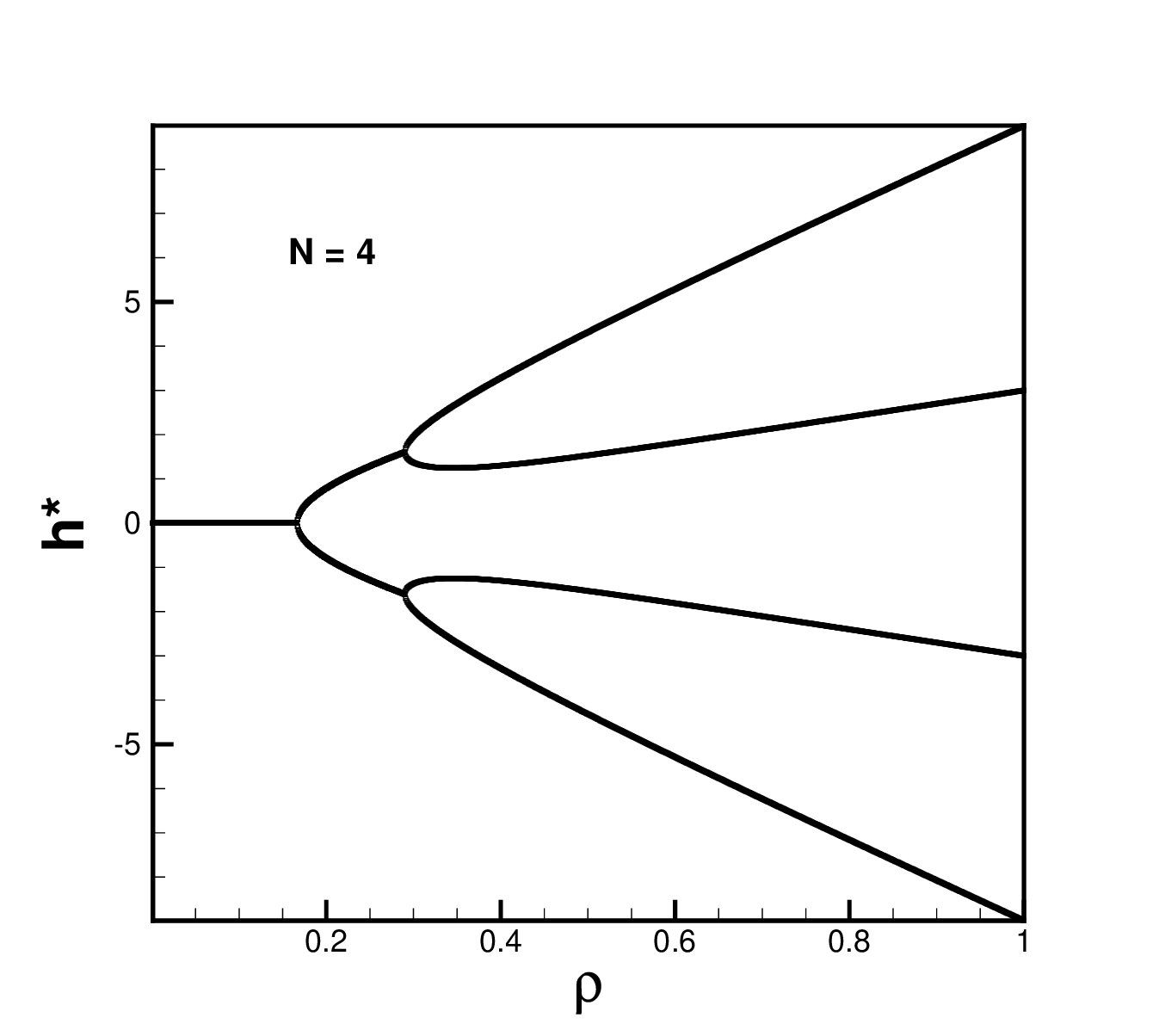}
\caption{Bifurcations of stationary values of fixed point components
(system \ref{Necuaciones}) depending on $\rho$, for the symmetric
case of reference when $N=4$.} \label{N4}
\end{figure}

\begin{figure}[t]
\centering
\includegraphics[width=0.45\textwidth]{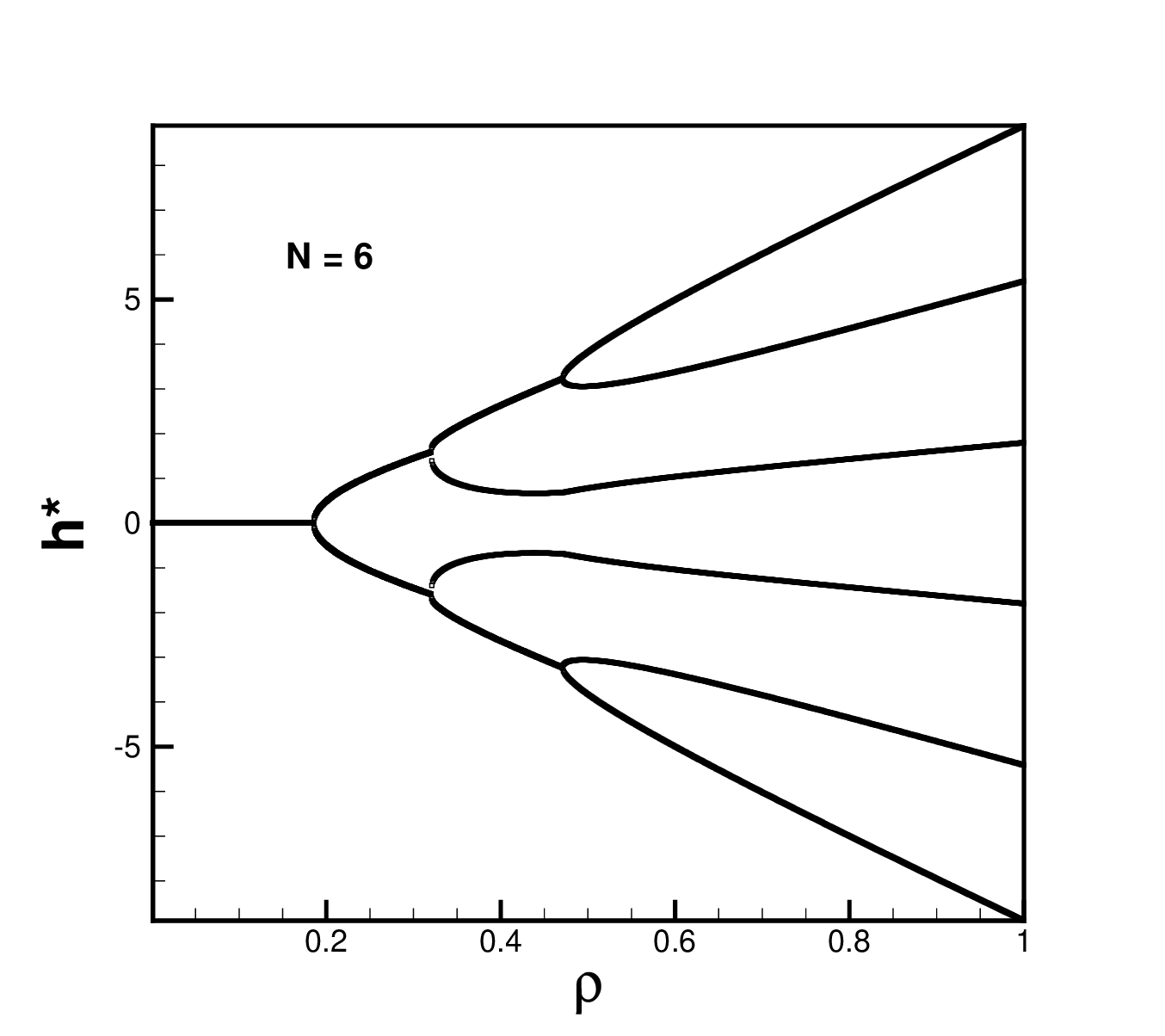}
\caption{Bifurcations of stationary values of fixed point components
(system \ref{Necuaciones}) depending on $\rho$, for the symmetric
case of reference when $N=6$.} \label{N6}
\end{figure}

\begin{figure}[t]
\centering
\includegraphics[width=0.45\textwidth]{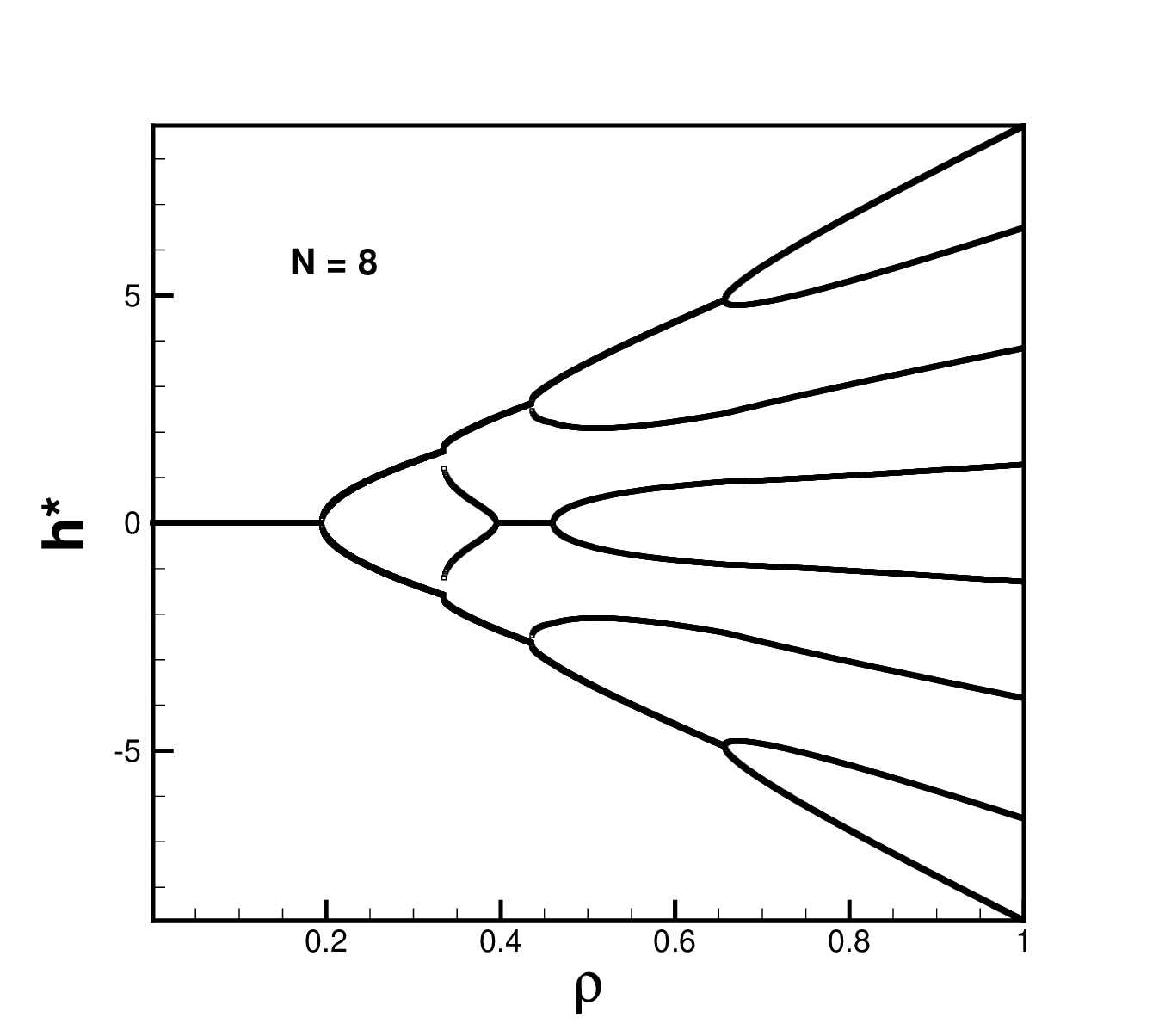}
\caption{Bifurcations of stationary values of fixed point components
(system \ref{Necuaciones}) depending on $\rho$, for the symmetric
case of reference when $N=8$.} \label{N8}
\end{figure}

\begin{figure}[t]
\centering
\includegraphics[width=0.45\textwidth]{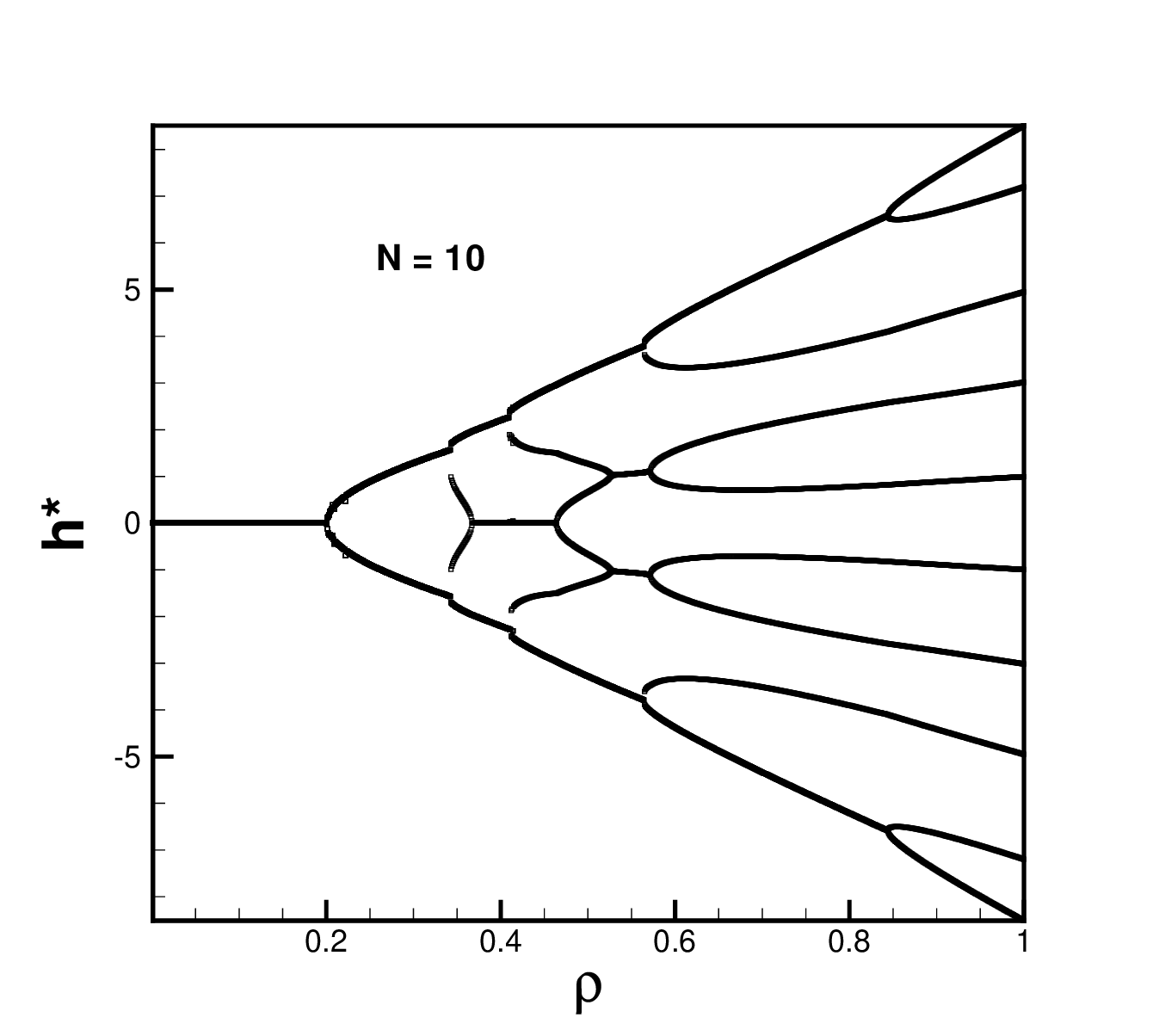}
\caption{Bifurcations of stationary values of fixed point components
(system \ref{Necuaciones}) depending on $\rho$, for the symmetric
case of reference when $N=10$.} \label{N10}
\end{figure}

Comparing additive to multiplicative relaxation in the Bonabeau
model, we must say that additive works worse than multiplicative:
status differences grow excessively and then, computing limitations
are exceeded (due to the exponential
explosion).\\
Anyway, if we introduce a new parameter $Q\geq0$ (instead of
increasing by one the winner status, we increase it by $Q$, so that
we can tune both winning and losing effects $Q,F$), the system
doesn't explode numerically with a good tuning of the parameters. In
 figure \ref{additive} with $N=10$, we took $F=0.7$, $Q=0.7$,
$\mu=0.0001$ and $\eta =0.001$. As we see, status reach values of
five order of magnitude, what we think is not realistic.\\
The hierarchy scheme in the multiplicative relaxation model (figures
\ref{N3}-\ref{N10}) develops much more complexity than the additive
one. An increase of density, above critical one, stills generates
hierarchy, as common sense would have dictated us. This fact is
traduced by a periodic fixed point coordinates $h^*_i$ splitting,
even after the phase transition. This hierarchy growing is not
trivial; something that has for sure been unnoticed for the moment.
In the additive relaxation model the hierarchical structure is
simple, it doesn't change with $\rho$ at hierarchical phase. Instead
of that, there is a fixed point coordinates $h^*_i$ splitting at
$\rho_c$, and dynamical evolution in hierarchical phase is poor.



\section{Mean field in the Stauffer version}

\noindent The model developped by Stauffer et al. comes from
Bonabeau's. It was firstly introduced to carry out the supposed lack
of transition of the previous model, discussed numerically in
\cite{Stauffer1,Stauffer2,Stauffer5}. In this version, the free
parameter $\eta$ is now exchanged with the order parameter $\sigma$,
such that
\newline
\begin{equation}
P_{ij}(t)=\frac{1}{1+\exp\big(\sigma(t)(h_j(t)-h_i(t))\big)}.
\end{equation}
 This modification somehow introduces a dynamical feedback to the
system: probability of winning/loosing is directly related to the
global inequality of the system, therefore, depending on the
climate's aggressiveness of the system, agents will behave more or
less aggressive themselves.

Just as in the case of Bonabeau's model, by introducing a mean field
we expect to find and reproduce analytically the phase transition
that is
proclaimed in the literature.\\\\

\noindent In the case of two-automata system, we develop the mean
field equations having in mind that:\\

\noindent(\textbf{$1$}) Each automaton updates at each time step
with the same dynamics that the Bonabeau model. The probability
calculation is
different though ($\eta\mapsto\sigma$).\\
\noindent(\textbf{$2$}) At each time step, variable $\sigma$ is
updated: the order parameter is now a dynamical parameter of the
system and
therefore evolves with it.\\

\noindent We once again redefine  $h=h_2-h_1$. With this change of
variable, we pass from a three equation system ($h_1,h_2,\sigma$) to
a two equation system ($h,\sigma$), without lack of generality,
because $h_1$ and $h_2$ are related through the mean status
(system's energy). The system equation is therefore:\\
\begin{eqnarray}
&&h(t+1)=(1-\mu)h(t)+\rho(1-\mu)(1+F)\cdot\bigg(1-\frac{2}{1+\exp\big(\sigma(t)h(t)\big)}\bigg)\nonumber \\
&&\sigma(t+1)=\mid\frac{1}{1+\exp\big(\sigma(t)h(t)\big)}-\frac{1}{2}\mid
.\nonumber
\end{eqnarray}
\normalsize  It is likely to expect the same qualitative results of
the stability analysis of this system than those about Bonabeau's
model, that is to say, loss of stability of the egalitarian regime
(i.e. the fixed point $h^*(=h^*_2-h^*_1)=0, \sigma^*=0$) would
become unstable at
some critical density $\rho_c$).\\

\begin{figure}[t]
\centering
\includegraphics[width=0.45\textwidth]{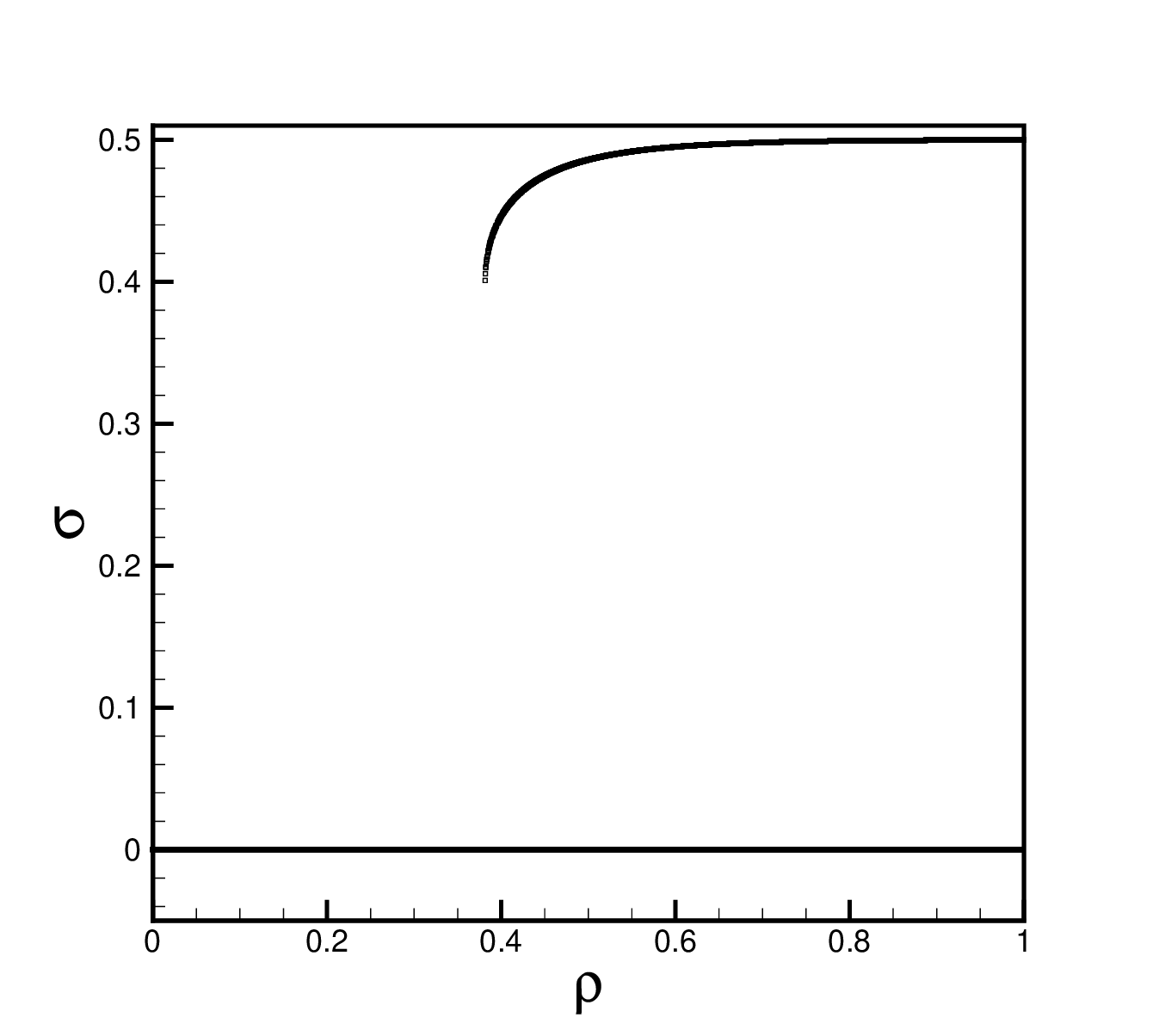}
\caption{$\sigma$ vs $\rho$ in the $2$ automata system. Initial
conditions for each $\rho$ are taken randomly. This leads to
stationary values of $\sigma$ being zero (egalitarian zone) or non
zero (hierarchical zone) depending of those chosen initial
conditions.} \label{Stauffer_ro_s}
\end{figure}

\begin{figure}[t]
\centering
\includegraphics[width=0.45\textwidth]{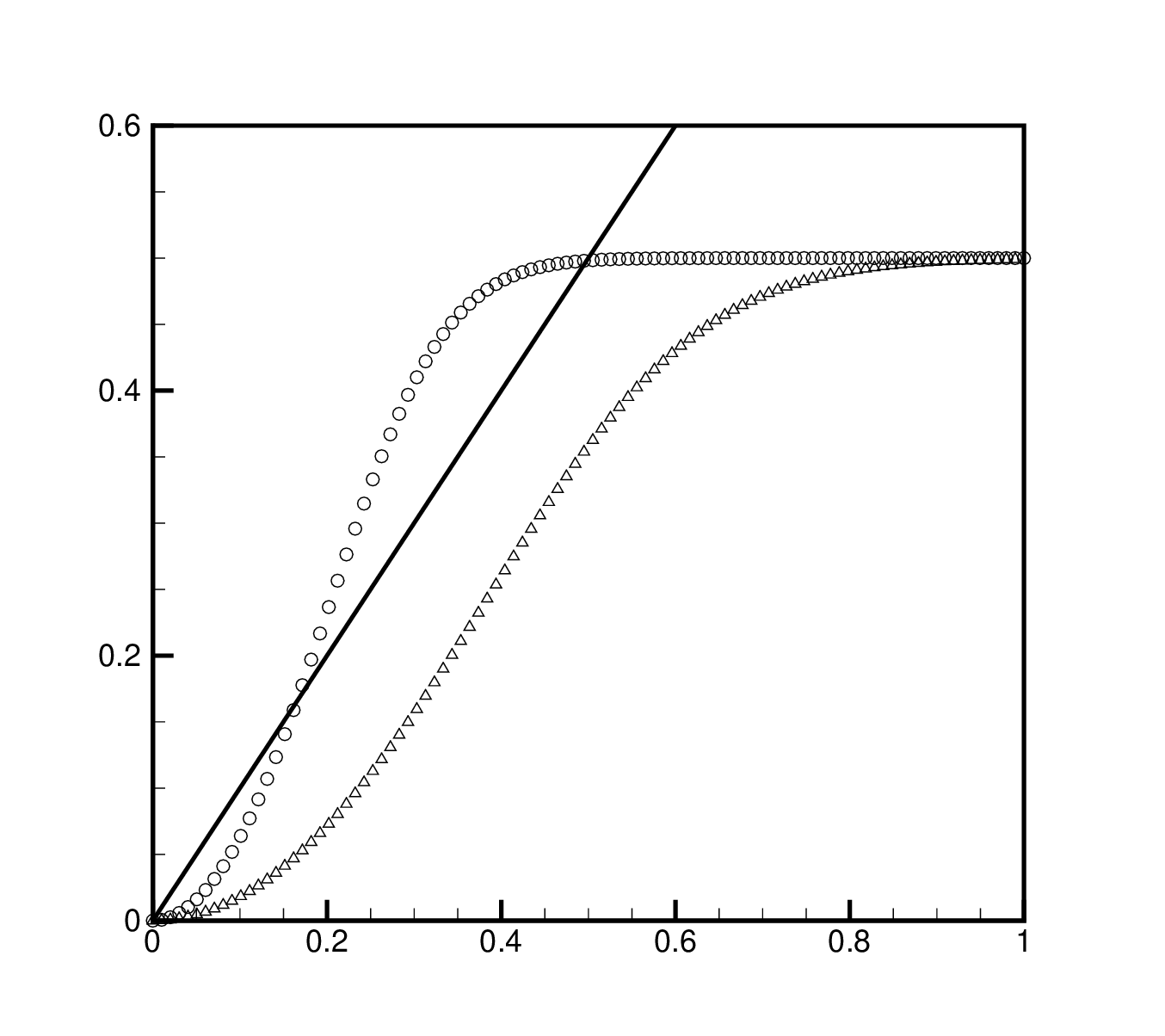}
\caption{Crossover from one stable fixed point of the system at
$\rho<\rho_c$ (triangles) to two stable fixed points
-egalitarian/hierarchic- (circles) and an unstable fixed point as
the delimiting branch between both domains of attraction.}
\label{stauffer_PF}
\end{figure}

The  Jacobian matrix of the linearized system is, in general:
\newline
$$J = \left(
            \begin{array}{cc}
              1-\mu+AB\sigma^* & 2ABh^* \\
              -\sigma^*A & -h^*A \\
            \end{array}
          \right),$$
\newline
where:
\newline
\newline
$A=\exp(\sigma^*h^*)/\big(1+\exp(\sigma^*h^*)\big)^2$,
\newline
\newline
$B=\rho(1-\mu)(1+F)$.
\newline
\newline
Evaluating $J$ in the egalitarian fixed point ($h^*=0; \sigma^*=0$),
we find that eigenvalues of $J$ are:
\newline
\newline
\begin{equation}
\lambda_1=0 < 1  ,   \lambda_2=1-\mu <1 , \forall \rho
\end{equation}
We find that egalitarian zone is stable for all densities. How come
a phase transition can then occur, as is presented in many previous
model simulations
\cite{Stauffer1,Stauffer2,Stauffer3,Stauffer4,Stauffer5,Stauffer6,Gallos}?
How come hierarchical situation can be achieved starting from
equality, if the egalitarian zone is always stable? The key of the
dilemma is set on the simulation methods that have been applied
until now. Figure (\ref{Stauffer_ro_s}) shows stationary values of
order parameter $\sigma$ $vs.$ control parameter $\rho$. For each
$\rho$, we take random initial conditions for status and $\sigma$
(these can be both zero or non-zero). The figure shows stationary
value $\sigma=0$ for all initial conditions, below a certain
density. Above it, we have stationary values of $\sigma$ being zero
in some cases (we remain in the egalitarian zone) and non-zero in
others (hierarchical
zone).\\
Numerically we observe that the system has one stable fixed point
below the critical density $\rho_c$ ($h^*=0; \sigma^*=0$) , and two
stable fixed points above ($h^*=0; \sigma^*=0$, $h^*\neq0;
\sigma\neq0$). The egalitarian zone is therefore always stable
($\forall \rho$). At $\rho_c$, a saddle-node bifurcation occurs, and
brings about the hierarchic branch. Notice that the stability scheme
is totally different from what we founded in the Bonabeau model:
while in that model, equality-hierarchy transition was generated
across Pitchfork bifurcation, due to loss of stability of the
egalitarian regime, in the Stauffer version the egalitarian regime
is always stable, but here at $\rho_c$ a saddle-node bifurcation
takes place. The stable branch of this bifurcation is related to the
hierarchic regime, and the unstable branch (not drawn in figure
\ref{Stauffer_ro_s}) plays the role of
frontier between the two domains of attraction.\\
Depending  what initial conditions we give to ($h_i, \sigma$), (i.e.
depending in which domain of attraction we start), the system, above
$\rho_c$, will evolve towards the egalitarian domain or the
hierarchic one.\\
In figure \ref{stauffer_PF} we can understand how this stability is
developed. Below $\rho_c$ the system has only one fixed point
(triangles), indeed stable due to (13): every set of initial
conditions $h,\sigma$ will evolve towards $h^*=0, \sigma^*=0$
(egalitarian zone). Above $\rho_c$ the system has three fixed points
(circles), moreover, from (13) and figure \ref{Stauffer_ro_s} we
know that upper and lower fixed point are stable and characterize
both egalitarian and hierarchic zones, this leads to an unstable
fixed point between them, performing the frontier. If the initial
conditions belong to the egalitarian domain, the system will evolve
towards an asymptotic egalitarian state. On the contrary, if the
initial conditions belong to the hierarchic domain, the system will
evolve to an asymptotic hierarchical state.\\\\
If in the model simulations, we set egalitarian initial conditions,
the result would be the "absence of transition". But if we fix some
other initial conditions, this could lead us to interpret the
results as "existence of transition". In simulations made by
Stauffer \cite{Stauffer6} they say "for the first ten Monte Carlo
steps per site, $\sigma=1$ to allow a buildup of hierarchies". This
fact probably allows initial fluctuations develop so that initial
conditions will be in the hierarchic domain.

\section{Conclusions}
The Bonabeau model has been criticized
\cite{Stauffer1,Stauffer2,Stauffer5} in the last years. In this
paper we revisit Bonabeau model in order to obtain analytical
evidences that give clear proof of the phase transition that the
system shows. We obtain, in the model with multiplicative
relaxation, a high complex structure of the hierarchical regime, a
fact that we think deserves an in-depth
investigation.\\
The Stauffer version, which is an alternative to Bonabeau model,
proposed by Stauffer et al. \cite{Stauffer1,Stauffer2,Stauffer5}, is
the base of recent works, basically focused on simulations
\cite{Stauffer3,Stauffer4,Gallos}. In this paper we tackle this
version with the same philosophy applied in the Bonabeau model.
Surprisingly, this one doesn't show a phase transition in rigor, as
far as there is no sudden growth of hierarchy if we start
from equality.\\

\section{Acknowledgments} The authors thank M. Cordero, J. Olarrea
and I. Parra for valuable discussions. This work has been supported
by CICYT of the Spanish Government under Project no. $BFM2002-01812$
(BL).




\end{document}